# PERFORMANCE EVALUATION OF BEACON-ENABLED IEEE 802.15.4 UNDER NS2

Faiza CHARFI[1] and Mohamed BOUYAHI

[1]Laboratoire d'Electronique et de Technologies de l'Information
Ecole Nationale d'Ingénieurs de Sfax, 3038 Sfax, Tunisie
`faiza.charfi@enis.rnu.tn faiza.charfi@gmail.com`

## ABSTRACT

*The increasing demand for real-time applications has made the Quality of Service (Qos) support for wireless sensor networks (WSN) a fairly new research framework. In this paper, we propose an extended model of the Beacon enabled IEEE 802.15.4 including the Guaranteed Time Slot GTS allocation mechanism in the aim to analyze and evaluate network performances. Series of extensive simulations were performed to analyze the impact of the Beacon Order BO and the Superframe Order SO on the network performance based on commonly known metrics. In particular, we examine data packet delivery performance and the throughput for different duty cycle rates. Also, we analyze the impact of the number of nodes on collision probability. Thus, for high number of nodes, collision becomes higher and the reachability begins to decline slightly. We discuss and compare simulation results conducted under various parameter settings to the IEEE 802.11network.*

## 1. INTRODUCTION

With the rapid growth in wireless technologies, Wireless Sensor Networks (WSN) have become a significant research challenge, attracting research communities and industry engineers [8]. They are used in an increasing number of applications like health-care, environmental monitoring and home surveillance. The WSNs are intended to support time-critical applications which are an important class of services supported by the IEEE 802.15.4 standard. Control, actuation and monitoring are all examples of applications where the information must be delivered within some deadline.

The IEEE 802.15.4 is a standard for short range, low rate-bit and low cost wireless personal area networks. It provides MAC and PHY layers for ZigBee. The IEEE 802.15.4 MAC standard specification describes the individual node behaviour. To support time-critical applications, IEEE 802.15.4 offers a Guaranteed Time Slot GTS allocation mechanism at the network coordinator. The packets are transmitted on a superframe basis. Each superframe is divided into Contention Access Period CAP, where nodes contend among each other to send packets, and a Contention Free Period CFP, where nodes have GTSs to send packets without contention. The GTS allocation provides communication services to time critical data. It makes guarantees on packets delivery and delivery times to be transmitted to the network coordinator [12].

Several works present analytical models of an IEEE 802.15.4 network to evaluate and analyze the performance of IEEE 802.15.4 [13] [14]. They characterize system delay, throughput, frame drop rate, energy consumption, and compare the performance with the GTS traffic. In this work, we focused on the modification of the IEEE 802.15.4 module into the NS2 simulator, including additional settings, and its performance evaluation. The proposed 802.15.4 Beacon-enabled PAN model uses a slotted CSMA/CA algorithm with GTS mechanism. We study various scenarios that arise when the nodes interact. We focus on the impact of the IEEE 802.15.4 standard parameters, specially the Beacon and Superframe orders, on packet delivery ratio, throughput and collision [6].

The paper is organized as follows: section 2 gives an overview on the IEEE 802.15.4 standard and the GTS allocation mechanism. In section 3 we discuss related work and the motivation of this paper. Section 4 presents the IEEE 802.15.4 slotted CSMA/CA algorithm. Section 5 describes both 802.15.4 MAC and PHY primitives and outlines their drawbacks. Section 6 presents the details of the experimental setup. Section 7 outlines the implementation of some primitives used by the MAC and PHY layers. A discussion on simulation results follows in section 8. Section 9 highlights the difference between ZigBee and WiFi networks. Finally, section 10 concludes the paper.

## 2. OVERVIEW ON IEEE 802.15.4

This section provides a brief overview of the IEEE 802.15.4 focusing on the relevant standard parameters to this study. The 802.15.4 is a part of the IEEE family of standards for physical and link-layers for Wireless Personal Area Networks WPANs. The IEEE 802.15.4 physical layer offers a total of 27 channels, one in the 868MHz band, ten in the 915MHz band, and finally sixteen in the 2.4GHz band [1]. The raw bit rates on these three frequency bands are respectively 20 kbps, 40 kbps and 250 kbps [11][18].

The IEEE 802.15.4 can operate either in a Beacon enabled or a non-Beacon enabled mode. The non Beacon enabled mode is useful for light traffic between the network nodes. The channel access and contention are performed using an unslotted CSMA-CA mechanism. In a Beacon-enabled network, the coordinator sends periodic Beacons containing information that allows network nodes to synchronise their communications, and information on the data pending for the different network nodes [7]. In this mode, the nodes communicate over the network through a superframe structure Figure 1. Each superframe has an active period, during which nodes can attempt to communicate using slotted CSMA/CA, and an inactive period during which devices may turn off in order to conserve energy. The active period is composed of three parts: a Beacon, a contention access period CAP and a contention free period CFP [12].

The IEEE 802.15.4 specification defines the Beacon, the MAC control, data and acknowledgment frames. All frames use a slotted CSMA/CA mechanism to access the channel except acknowledgment frames and data frame that follows the acknowledgment of a data request command, which are transmitted in the CAP.

The CFP is used for GTS allocation, which can support Qos for real time applications. A node can request the PAN coordinator to allocate GTS for low latency applications. In turn, the PAN coordinator can allocate at most seven GTSs [9]. During the GTS, nodes are allowed to transmit without any contention with other devices. Figure 1 shows a request for GTS deallocation and an update of CFP [11].

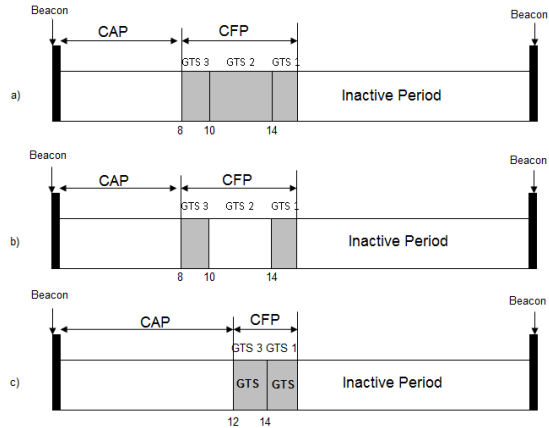

Figure 1. Superframe structure with GTS reservation examples [11].

## 3. RELATED WORK

Some work has been conducted in evaluating the performance of the 802.15.4 standard. In [4], the authors evoked a Beacon-enabled transmission in star networks. They study the effective compromises between power consumption and throughput or latency. They prove that small node's duty cycle can give substantial energy savings. Also, the energy cost of synchronizing to the Beacons is significant. Author in [2] underline the suitability of 802.15.4 in wireless medical applications used in patient care applications through the network performance evaluation. Their focus is on interoperability and scalability. The study in [6] concentrates on power consumption. They determine the minimum expected power consumption in a typical WSN scenario and examine how energy is used in different phases of data transmission.

J. Zheng and M.J. Lee in [6] implemented the IEEE 802.15.4 standard on NS2 simulator and provided simulation-based performance evaluation on 802.15.4. It was a comprehensive literature that defines the 802.15.4 protocol and was mainly confined to the IEEE 802.15.4 MAC performances. This work has a minor evaluation on the performance of the peer-to-peer networks [7]. Our work here focuses on simple 1-hop star network. It describes the wireless sensor networks in the IEEE802.15.4 to integrate the GTS mechanism in the MAC layer in order to improve the Qos. To achieve this work, we chose the NS2 simulator. The Focus is on extending this simulation paradigm by introducing additional settings and performance metrics.

## 4. THE CSMA/CA ALGORITHM

The CSMA-CA is used for the devices to communicate with each other. It implements Backoff periods BP which is defined as aUnitBackoffPeriod. The slotted CSMA/CA Backoff algorithm uses three variables which are the Number of Backoff NB, the Contention Window CW and the Backoff Exponent BE. The NB defines the number of times the CSMA/CA algorithm needs to access the channel while attempting the current transmission. CW represents the number of Backoff periods required to be clear of channel activity before starting transmission. BE enables the computation of the Backoff delay to assess the channel and to reduce the collision probability [11]. In reference to the standard, the variables NB, CW and BE are initialized respectively to 0, 2, min (2, macMinBE) depending on the value of the Battery Life Extension (first step). BE must not exceed aMaxBE which is by default set to 5, whereas macMinBE has a default value of 3. The MAC layer waits for random unit of Backoff periods within 0 to $2^{BE} - 1$ before sensing the channel (second step) Figure 2. As the Backoff is expired, the PHY protocol performs Clear Channel Assessment CCA on the channel to detect any activities and relays the results to the MAC layer (third step). The MAC sublayer attempts the frame transmission in the CAP duration, otherwise it waits for the next superframe's CAP and repeats the evaluation. If the channel is busy, the MAC sublayer reinitializes the CW to 2 Backoff slots and increments both NB and BE by one (fourth step). If the NB value exceeds macMaxCSMABackoffs which is set to 5, CSMA-CA terminates with a failure status, otherwise it returns to the second step [3]. If the channel is found to be idle, the MAC sublayer decrements by one the CW (fifth step). If the latter is different from 0, the CSMA-CA returns to the third step. Otherwise, it begins the transmission in the next Backoff period. In the unslotted CSMA-CA, the transmission starts immediately if the channel is found to be idle [17].

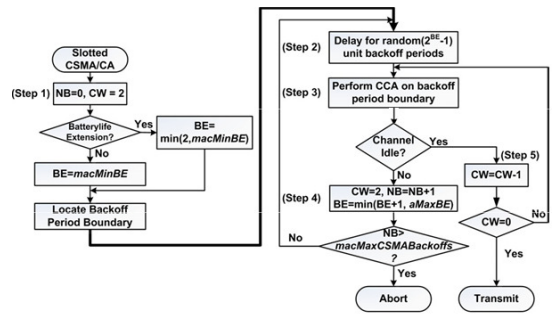

Figure 2. IEEE 802.15.4 Slotted CSMA/CA.

## 5. PRIMITIVE MAC AND PHY LAYERS OF IEEE 802.15.4

The services of a layer define the offered capabilities to ensure the information flow between the users and the layers. This information flow is modeled by discrete events, which characterizes the provision of a service primitive. The IEEE 802.15.4 standard specifies 14 PHY and 35 MAC primitives and supports two types of devices, the FFD and the RFD [8]. The FFD is a full function device supporting all the defined primitives of the standard whereas the RFD is a reduced function device. Some of these primitives are described in brief here to better understand the software implementation of the Zigbee modules.

### 5.1 PHY Layer primitives

The physical layer is an essential component in computer communication. It is responsible for the activation and deactivation of the radio transceiver, energy detection, link quality indication measurement, clear channel assessment, data transmission and reception. It interacts directly with the wireless channel supplying information to and from the upper layers. The PHY primitives indicate the functions organized by each layer. Whenever there is data to be transmitted, the MAC Layer Management Entity MLME calls the PHY layer with two primitives PD-DATA.request and PD-DATA.confirm to transmit a data frame. For the data packet reception, the primitive PD-DATA.indication is generated by the PHY entity and issued to its MAC sublayer entity to transfer a received PHY Service Data Unit PSDU [11].

The Clear Channel Assessment CCA is implemented by two primitives, PLME-CCA.request and PLME-CCA.confirm to identify if the channel is free or busy. The CSMA-CA algorithm invokes the PLME-CCA.request whenever an assessment of the channel is required.

The function of activating and deactivating the transceiver is performed by the primitives PLME-SET-TRX-STATE.request and PLME-SET-TRX-STATE.confirm to change the operating state of the transceiver.

International Journal of Distributed and Parallel Systems (IJDPS) Vol.3, No.2, March 2012

### 5.2 MAC Layer primitives

The Mac layer provides an interface between upper layers and the PHY layer. It handles through its primitives various mechanisms such as Beacon Transmissions, synchronization to the Beacons, PAN Association/Disassociation, CSMA-CA for Channel Access and GTS transmissions.

Two primitives are used for data transmission from the higher layer. The MAC Common Part Sublayer MCPS-DATA.request requests the transmission of a data unit to the PHY layer. The result is indicated with the MCPS-DATA.confirm primitive as a response to successful data transmission. The MCPS-DATA.indication primitives are used for data reception from the PHY lower layer to indicate the transfer of data from the MAC sublayer to the recipient next higher layer [11].

Four types of primitives are used to provide association services: MLME-ASSOCIATE.request allows a device to request an association with a coordinator. MLME-ASSOCIATE.indication indicates the reception of an association request command; MLME-ASSOCIATE.response is used to initiate a response whereas the MLME-ASSOCIATE.confirm primitive is used to inform the initiating device of the successful or unsuccessful association.

The primitive scan is used for the detection of energy, active scan, passive scan and orphaning scan. Thus, the MAC layer performs the primitive PLME-ED.request for the energy detection. The device that has lost contact with its associated PAN sends the orphaning request for any channel for the PAN detection.

To provide GTS services three types of primitives are used: MLME-GTS.request allows a device to request a GTS to the coordinator. The GTS request is generated at the next higher layer; MLME-GTS.indication indicates the reception of a GTS request command and the MLME-GTS.confirm primitive is used to inform the initiating device of the successful or unsuccessful GTS [11].

## 6. EXPRIMENTAL SETUP

The NS2 version 2.26 was used for carrying out some simulation experiments, with patches for the IEEE 802.15.4 LR-WPAN code implemented by J.Zheng [6]. The implementation covered the essential functionalities except security and the contention free period which consisted of slot reservations for Qos application. In the current experiments, we adopted the same PHY layer and radio parameters. Some extensions to the Mac layer were introduced to accommodate to the proposed simulation settings. We considered the 2.4GHz frequency band due to its larger scale sensor deployment. A Beacon enabled star topology network is studied. It is assumed that the Beacon interval is composed of active and inactive parts. The simulation scenarios were run in static environment where n FFD nodes (varying from 5 to 25) were distributed around a circle of 10 meter radius, with the PAN coordinator at the center Figure 3. The decoding and the sensing range thresholds of the nodes were set to 18 meters, so that all nodes formed a single cell. We used the Constant Bit Rate CBR traffic for all simulation sessions [4][15][16][19].

7171

International Journal of Distributed and Parallel Systems (IJDPS) Vol.3, No.2, March 2012

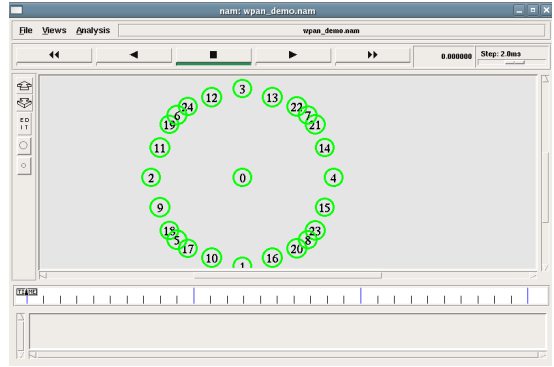

Figure 3. The used topology.

## 7. GTS IMPLEMENTATION

The current implementation of the superframe structure includes the following parameters of simulation: the superframe order (SO) which manages the duration of the active period (1), Beacon order (BO) which manages the duration of the superframe (2) and the slot duration (sd) which defines the slot duration (3). aBaseSlotDuration equals 60 symbols resulting 12.5 ms whereas aBaseSuperframeDuration equals 960 symbols.

$$SD = aBaseSuperframeDuration * 2^{SO} \qquad (1)$$

$$BI = aBaseSuperframeDuration * 2^{BO} \qquad (2)$$

$$sd = aBaseSlotDuration * 2^{SO} \qquad (3)$$

The slotted CSMA/CA MAC protocol of IEEE802.15.4 is now extended to include the GTS mechanism. Hence, we implement the primitive GTS according to the IEEE 802.15.4 protocol specification. In this section, the GTS primitive code and a screenshot are shown to verify how our implementation is working. To allocate slots in the CFP, the device shall send a request to its coordinator and wait for the GTS Descriptor in the Beacon payload [11]. In the Beacon, the GTS is enabled through the GTS field (Permit = 1) but there is still no slot allocated (Len=0). The device, namely node 1, sends a GTS request to the coordinator (node 0) at time 27,0016001ms Figure 4. The GTS length is 2; thus two slots are reserved and the direction is set to 1 for device reception. The new CAP length is defined by FinCAP which is equal to 13. Index 2 indicates the time of the reception of GTS request by the coordinator. The GTS characteristics are passed to the coordinator through the GTS request in the the MSDU_Payload[0] field. The GTS count is one by GTS Specification and corresponds to gtsSpec.count. The start slot list[0].slotSpec has fourteen positions and is returned in left quartet. Two slots are reserved and returned in the GTS list field right quartet by list[0].slotSpec. The index 3 corresponds to the Ack reception by the node 1; finally index 4 represents the reception of the frame Beacon by node 1.

Figure 4. Result of simulation of primitive GTS.

The pan descriptor sender within the coordinator (panDes2.list[0].devAddr) represents the address of node reserved. The panDes2.list[0].slotSpec corresponds to the start slot. The panDes2.list[0].dir represents the direction of transmission. panDes2.SuperframeSpec represents the superframe specification. panDes2.list[0].length designs the length of slot reserved.

The following code is an implementation of the GTS in the Service Specific Convergence Sublayer SSCS to support GTS allocation mechanism.

```
\$node_(1)    sscs    startGTSDevice    <GTSCharacteristics=0x1c>    <txOption=0x02>
<gtsPermit=1> <BeaconOrder=3> <SuperframeOrder=3>
channel scan
if coordinators not found
    association fails
elseif no coordinators permit association
    association fails
else
    select a proper coordinator
    send association request to the coord.
    wait for ACK
    if ACK not received
        association fails
    else
        send data request to the coord.
          wait for ACK
          if ACK not received
              association fails
          else
              wait for association response
              if asso. response not received
                  association fails
              elseif association not granted
        association fails
              else
        association succeeds
    send gts request to the coord.
    wait for ACK
    if ACK not received
        gts fails
```

```
else
    wait for BCN
    if BCN not received
        gts fails
    else
        gts succeeds
```

## 8. SIMULATION RESULTS

An extensive series of simulations was carried out using the IEEE 802.15.4 Beacon-enabled cluster with GTS mechanism. Additional settings (BO, SO) are included to investigate their effects on the network performances. After analysing the trace file, we extract the following metrics which are used to study the performance of the proposed model of the IEEE 802.15.4. All metrics are defined with respect to MAC sublayer and PHY layer in order to isolate their effects from those of upper layers. In this work, simulation results consider Beacon order value <6. Indeed, BO= 6 and above are not suitable for wireless sensor networks because they delay excessively the node's association time.

### 8.1 Throughput by node

The first experiment considers 10 nodes distributed around one coordinator. The aim is to study the throughput when different Beacon orders and different duty cycles are being set up. The throughput is calculated from the ratio of total bytes received to total time of simulation multiplied by the nodes number multiplied by one thousand (4).

$$S = \frac{total\_received\_bytes * 8}{simulation\_time * number\_nodes * 1000} \quad (4)$$

In this section, we investigate the impact of beacon order on the throughput. We assume the allocation of only one time slot GTS in each super-frame. In what follows, the change of the SO means that the beacon order also changes satisfying SO = BO. Similarly, we have made the simulation with the BO value ranging 1 from 6 Figure 5.

We remark, in the upper curve, that the closer the values of SO and BO, the more significant the throughput. However, as the inactive part increases, the throughput drops significantly. Based on the results obtained, we conclude that large inactive period can decrease throughput performance which is influenced by processing, transmitting, propagation, and queuing delays.

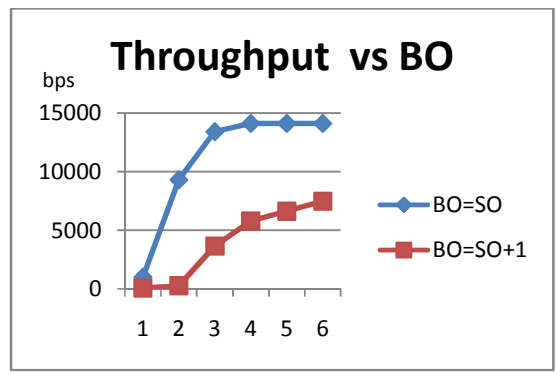

Figure 5. Throughput Vs BO.

As the energy savings in the beacon-enabled mode depend on the amount of periodic sleep periods introduced, it is important to control the fraction of the time that the node is active. This time, known as duty-cycle, is computed as the ratio between the superframe duration and the beacon interval that can be related to BO, SO.

Figure 6 presents the throughput performance against varying the duty cycle by fixing BO and adapting SO to traffic. We observed that, if SO is fixed to a low value, the WSN produce lower network throughput than a higher SO value with the same duty-cycle ratio. Also, when the value of 100% duty cycle is incremented, the active period of the superframe becomes longer, which causes the increase of the transmitted packet. The scenario becomes worse with the case of inactivity period between the Beacons which result on increased amount of wasted bandwidth of an allocated GTS.

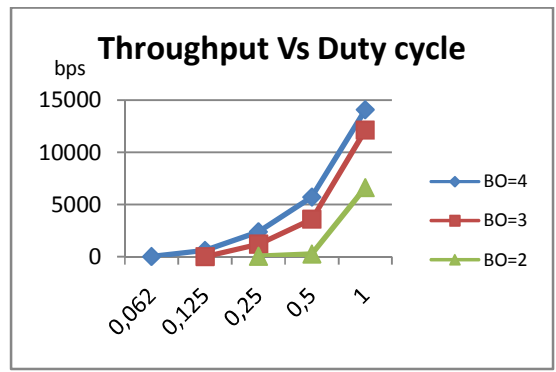

Figure 6. Throughput Vs Duty Cycle.

## 8.2 Packet Delivery Ratio

The second experiment evaluates the Packet Delivery Ratio which is defined by the ratio of packets successfully received to packets sent in MAC sublayer (5).

$$Pd = \frac{total\_received\_packets}{total\_sent\_packets} *100 \qquad (5)$$

This metric does not differentiate transmissions and retransmissions, and therefore does not reflect what percentage of upper layer payload is successfully delivered, although they are related. Results from figure 7 show that lower Beacon order values such as 0, 1, 2 or 3 decrease the Packet Delivery. This is due to the fact that the node orphans too frequently, and is busy associating and re-associating itself rather than receiving data. The results become stable for Beacon order 4 and above.

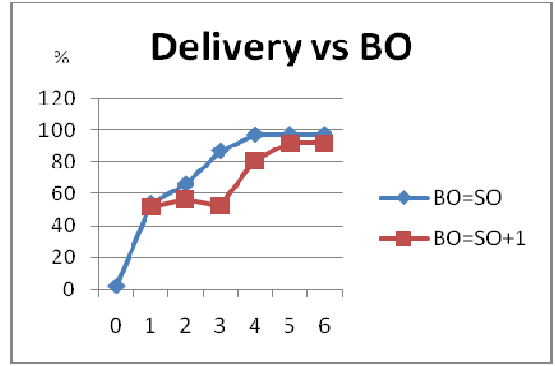

Figure 7. Delivery Vs BO.

The scenario becomes worse with the case of inactivity period between the Beacons and the packets delivery ratio decreases notably Figure 8.

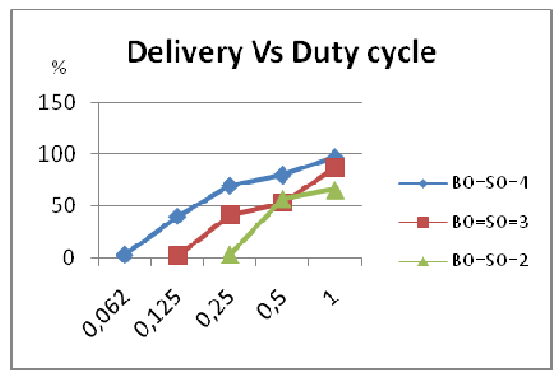

Figure 8. Delivery Vs Duty Cycle.

## 8.3 Collision rate between terminals

Experiment 3 consists in investigating the collision rate between terminals. Thus the number of nodes is varied from 5 to 15. Also, we consider only the active period and the Beacon order (BO) is varied from 0 to 6.

The total collisions that occur between terminals during a simulation run are defined by the ratio of all the packets of data removed to all the removed packets multiplied per hundred (6) [6].

$$C = \frac{total\_data\_collision\_packets}{total\_collision\_packets} *100 \qquad (6)$$

As the number of node grows, the percentage of the removed packets increases. Thus, the large number of nodes will introduce a high level of collisions which decreases with the increase of the active period Figure 9. Therefore, we deduce that the collision percentage is proportional to the number of nodes and inversely proportional to the duration of the active period.

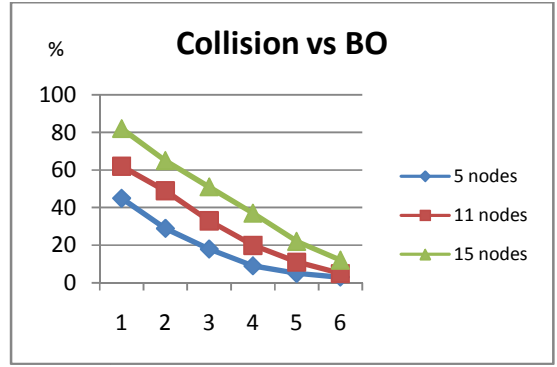

Figure 9. Collision Vs BO.

## 9. COMPARISON NETWORKS ZIGBEE AND WIFI

This section compares the Beacon enabled IEEE 802.15.4 (Zigbee) and the IEEE 802.11 (Wifi) standards in regard to the previous performance metrics for wireless networks: the throughput by node and the Packet Delivery Ratio.

### 9.1 Throughput by node

Figure 10 shows that the throughput by node in the case of Wifi is better than ZigBee because it provides significantly higher aggregate bandwidth. It is known that the Wifi network does not exploit all the band-width, but in the case of ZigBee the band-width is limited. Additionally, the drop in throughput can be seen in both networks as the number of nodes rises. Thus, the growth of the nodes number can seriously degrade the network throughput by loss of node synchronization which is due to the increase in the number of Beacons lost in the case of Zigbee and packet collisions in the case of Wifi and also Zigbee.

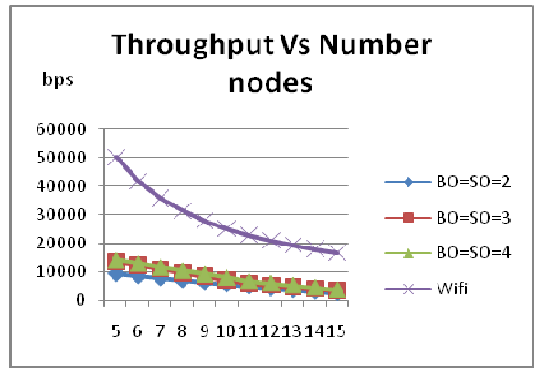

Figure 10. Throughput Vs Number Nodes.

### 9.2 Packet Delivery Ratio

The IEEE 802.11/Wifi offers high Packet Delivery Ratio of 99% Figure 11. By contrast, the Beacon enabled IEEE 802.15.4 results in high Delivery Ratio close to that of WiFi while active transmitting. Indeed, the values of BO=5 provides a Packet Delivery Ratio of 95% and sound better than the Beacon order values 2 and 4.

International Journal of Distributed and Parallel Systems (IJDPS) Vol.3, No.2, March 2012

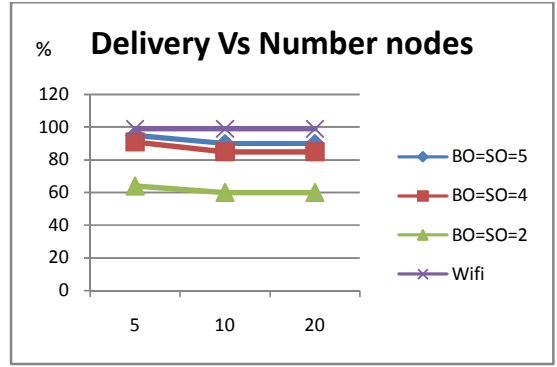

Figure 11. Delivery Vs Number Nodes.

## 10. CONCLUSION

In this paper, we focused on the modification of the IEEE 802.15.4 module into the NS2 simulator and its performance evaluation. We have extended this simulation model about Guaranteed Time Slot (GTS) mechanism supporting deterministic real-time traffic. Additional settings (BO, SO) are included to investigate their effects on the network performances. Hence, we analyzed the achieved throughput for different Beacon order values versus the number of nodes and the duty cycle. We observed that lower Beacon order gives a worse throughput because of the higher packet drop probability. Also, the growth of node number with higher Beacon order increases significantly the collision rate between the terminals and degrades the throughput due to wasted bandwidth. The IEEE 802.15.4/Zigbee using the Beacon mode achieves high Packet Delivery Ratio close to that of Wifi while active transmitting and standby period can be adjusted. Our future works will further investigate solutions to improve the performances of IEEE 802.15.4 for large scale Wireless Sensor Networks.